\documentclass[prd,reprint,showpacs,showkeys]{revtex4-1}
\usepackage{amsfonts}
\usepackage{amssymb}
\usepackage{amsmath}
\usepackage{graphicx}
\usepackage[font={footnotesize,it}]{caption}

\begin{document}

\title{Counterrotational effects on stability of 2+1-dimensional thin-shell
wormholes}
\author{S. Habib Mazharimousavi}
\email{habib.mazhari@emu.edu.tr}
\author{M. Halilsoy}
\email{mustafa.halilsoy@emu.edu.tr}
\affiliation{Department of Physics, Eastern Mediterranean University, Gazimagusa, north
Cyprus, Mersin 10, Turkey. }
\date{\today }

\begin{abstract}
The role of angular momentum in a $2+1$-dimensional rotating thin-shell
wormhole (TSW) is considered. Particular emphasis is made on stability when
the shells (rings) are counterrotating. We find that counter-rotating halves
make the TSW supported by the equation of state of a linear gas more stable.
Under a small velocity dependent perturbation, however, it becomes
unstable.\qquad
\end{abstract}

\maketitle

\section{Introduction}

Similar to the black holes, wormholes in $2+1-$dimensions also constitute
informative objects to help us learn more about their higher dimensional
counterparts. Same is also true for the thin-shell wormholes (TSWs) \cite%
{MV1} which are constructed from an energy-momentum at the throat through
cut-and-paste technique satisfying the proper junction conditions. We recall
that the history of TSWs started with the Visser's construction in a $3+1-$%
dimensional flat Minkowski spacetime \cite{MV1}. At the same time he
extended the idea to the Schwarzschild spacetime \cite{MV2}. In his book 
\cite{MV3} and work with Poisson \cite{MV4} they introduced into physics the
terminology of TSWs first. In the latter work they considered also the
stability of a TSW. Ever since, there have been extensive attempts to
understand different features of the new kind of wormholes. For a short list
of such attempts we refer the readers to \cite{TSW} and references cited
therein. Among the works on TSW we see generalizations to higher \cite{HD},
lower \cite{LD} dimensions and to the extend of Lovelock gravity \cite%
{LOVELOCK}. The theory has also been considered in cylindrical symmetry \cite%
{CYLIN}, Dilaton theory \cite{Dilaton} and so on. The common feature of all
these extensions is that the bulk spacetimes are all static. We are
well-aware about the difficulties in rotating thin-shell wormholes (RTSW).
For this reason we restrict ourselves in this study to the relatively
simpler case of $2+1-$dimensions. The absence of gravitational degrees of
freedom, namely, Weyl curvature in the lower dimension enforces us to add
new physical parameters such as cosmological constant, electric / magnetic
charge, scalar charge and rotation. Our aim in this study is to consider
RTSWs in $2+1-$dimensions. The metric function contains the square of
angular momentum $\mathbf{J}^{2}$ which doesn't distinguish the cases of $%
\pm \mathbf{J}$ for angular momentum. We recall that rotating TSWs
constructed from Kerr black hole in $3+1-$dimensions have been considered in 
\cite{Kerr} and their geodesics have been analyzed in \cite{Kerr2}. Also
rotational effects for collapsing thin-shells in $2+1-$ and $4+1-$dimensions
have been considered in detail in \cite{MANN} and \cite{Delsate}
respectively.

In our analysis of TSWs we observe that the off-diagonal components of the
extrinsic curvature tensor $k_{ij}$ and related components of the surface
energy-momentum at the throat $S_{ij}$ vanish in case we assume
counterrotating components of shells at the throat. The gas pressures from
the upper and lower shells cancel each other to modify the equation of state
to the extend that it becomes equivalent to a static case. We may draw a
rough analogy from the rotating Earth. Due to the non-inertial effects curly
geodesics of winds i.e. the Coriolis effect matching at the equator are
counter-circular in different hemispheres. If the two hemispheres of our
Earth were counterrotating instead of corotating the curly motions should be
identical. Counterrotation at the throat in the case of TSWs allows us to
choose a simpler surface energy-momentum tensor and study the stability
condition. We can choose, for instance, a linear gas (LG) equation of state
(EoS) at the junction in which the pressure is linearly related to the mass
density. For such a LG the counterrotating components make the TSW more
stable. Increasing angular momentum magnitude i.e. the relative rotation
stabilizes the TSW further. Our next attempt toward a stable TSW is to
assume a velocity dependent perturbation of the LG. It is observed that our
stability argument is restricted by the linear perturbation alone in which
at the equilibrium radius ($R_{0}$) we have the initial conditions $\dot{R}%
_{0}=\ddot{R}_{0}=0.$ Once we assume that $\dot{R}_{0}\neq 0,$ the perturbed
throat grows exponentially to make the TSW unstable. This behavior is proved
explicitly.

\section{Rotating Thin-Shell Wormhole}

%%%%%%%%%%%%%%%%%%%%%%%%%%%%%%%%%%%%%%%%%%%%%%%%%%%%%%%%%%%%%%%%%%%%%%%%%
\begin{figure}[tbp]
\includegraphics[width=70mm,scale=0.7]{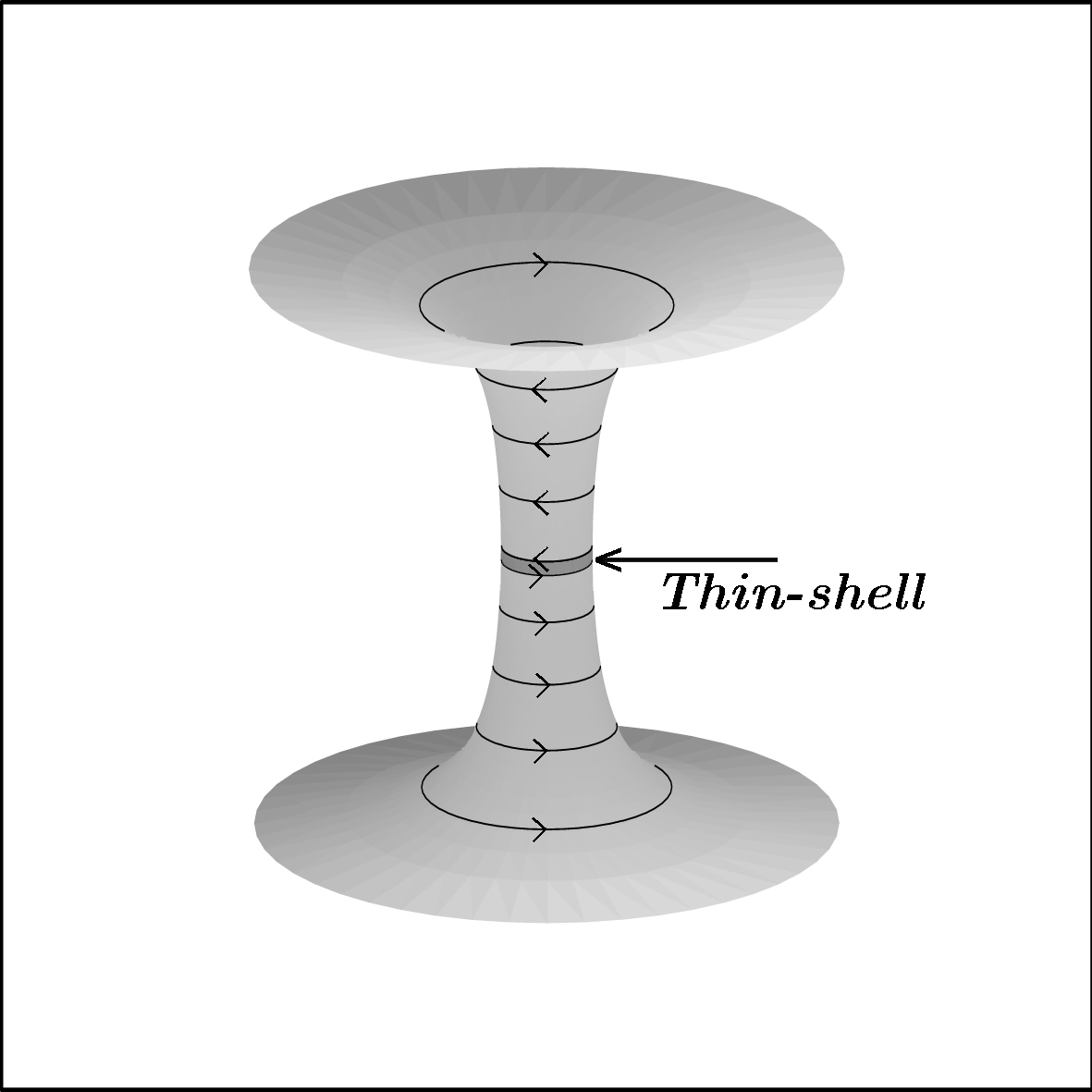}
\caption{Rotating thin-shell wormhole made by counterrotating perfect
fluids. }
\end{figure}
%%%%%%%%%%%%%%%%%%%%%%%%%%%%%%%%%%%%%%%%%%%%%%%%%%%%%%%%%%%%%%%%%%%%%%%%%%%%%

The $2+1-$dimensional rotating Ba\~{n}ados-Teitelboim-Zanelli (RBTZ) black
hole solution is \cite{BTZ}%
\begin{equation}
ds^{2}=-f(r)dt^{2}+\frac{dr^{2}}{f\left( r\right) }+r^{2}\left( d\varphi
+N^{\varphi }\left( r\right) dt\right) ^{2}
\end{equation}%
in which%
\begin{equation}
f\left( r\right) =-M+\frac{r^{2}}{\ell ^{2}}+\frac{J^{2}}{4r^{2}},
\end{equation}%
and%
\begin{equation}
N^{\varphi }\left( r\right) =-\frac{J}{2r^{2}}.
\end{equation}%
Herein $\Lambda =\pm 1/\ell ^{2}$ and $M$ and $J$ are the mass and angular
momentum of the RBTZ black hole. We note that in order to have two horizons $%
0\leq \frac{J}{\ell }\leq M$. To construct the thin shell wormhole we use
the method of cut-and-paste introduced in \cite{MV1,MV2,MV3,MV4,TSW}.
According to this method we take two copies of the bulk $\mathcal{M}_{\pm
}=\left\{ x^{\mu }|t\geq T\left( \tau \right) \text{ and }r\geq R\left( \tau
\right) \right\} $ with the line elements%
\begin{equation}
ds_{\pm }^{2}=-f_{\pm }(r)dt^{2}+\frac{dr^{2}}{f_{\pm }\left( r\right) }%
+r^{2}\left( d\varphi +N_{\pm }^{\varphi }\left( r\right) dt\right) ^{2}
\end{equation}%
and we paste them at an identical hypersurface $\Sigma _{\pm }=\Sigma
=\left\{ x^{\mu }|t=T\left( \tau \right) \text{ and }r=R\left( \tau \right)
\right\} .$ For convenience we move to a corotating frame by introducing $%
d\varphi +N_{\pm }^{\varphi }\left( R\right) dt=d\psi $ on each side
separately. The line elements in both sides become%
\begin{multline}
ds_{\pm }^{2}=-f_{\pm }(r)dt^{2}+ \\
\frac{dr^{2}}{f_{\pm }\left( r\right) }+r^{2}\left[ d\psi +\left( N_{\pm
}^{\varphi }\left( r\right) -N_{\pm }^{\varphi }\left( R\right) \right) dt%
\right] ^{2}.
\end{multline}%
The product manifold is complete with one boundary at the hypersurface $%
\Sigma $ which we shall call the throat. On the throat the line element is
given by%
\begin{equation}
ds_{\Sigma }^{2}=-d\tau ^{2}+R^{2}d\psi ^{2}.
\end{equation}%
To make the Israel junction conditions \cite{ISRAEL} satisfied, first of all
we have $f_{+}(R)=f_{-}(R)=f\left( R\right) $ followed by%
\begin{equation}
-f(R)\dot{T}^{2}+\frac{\dot{R}^{2}}{f\left( R\right) }=-1
\end{equation}%
in which a $dot$ stands for the derivative with respect to the proper time $%
\tau .$ Next, the Einstein's equations (Israel junction conditions) on the
hypersurface become%
\begin{equation}
k_{i}^{j}-k\delta _{i}^{j}=-8\pi GS_{i}^{j},
\end{equation}%
in which $k_{i}^{j}=K_{i}^{j\left( +\right) }-K_{i}^{j\left( -\right) },$ $%
k=tr\left( k_{i}^{j}\right) $ and 
\begin{equation}
K_{ij}^{\left( \pm \right) }=-n_{\gamma }^{\left( \pm \right) }\left( \frac{%
\partial ^{2}x^{\gamma }}{\partial X^{i}\partial X^{j}}+\Gamma _{\alpha
\beta }^{\gamma }\frac{\partial x^{\alpha }}{\partial X^{i}}\frac{\partial
x^{\beta }}{\partial X^{j}}\right) _{\Sigma }
\end{equation}%
is the extrinsic curvature with embedding coordinate $X^{i}$. Also the
normal unit vector is defined as 
\begin{equation}
n_{\gamma }^{\left( \pm \right) }=\pm \left( -\dot{R},\dot{T},0\right) .
\end{equation}%
The surface energy-momentum tensor of the throat is chosen to be a perfect
fluid type gas with%
\begin{equation*}
S_{ij}=\left( \sigma +P\right) u_{i}u_{j}+Pg_{ij}.
\end{equation*}%
Here $\sigma $ is the energy density of the shell, $P$ is the tangential
pressure at the throat and $u_{i}$ is the shell's velocity. We note that in
rotating system in both sides of the throat $u_{\psi }=0$ and therefore $%
S_{\tau \psi }=0$ which means that $S_{i}^{j}=$diag$\left( -\sigma ,P\right) 
$. The extrinsic curvature components are then found to be%
\begin{equation}
K_{\tau }^{\tau \left( \pm \right) }=\pm \frac{2\ddot{R}+f^{\prime }}{2\sqrt{%
\dot{R}^{2}+f}},
\end{equation}%
\begin{equation}
K_{\tau }^{\psi \left( \pm \right) }=\mp \frac{N_{\pm }^{\varphi }\left(
R\right) }{R}
\end{equation}%
and%
\begin{equation}
K_{\psi }^{\psi \left( \pm \right) }=\pm \frac{\sqrt{\dot{R}^{2}+f}}{R}.
\end{equation}%
As a result%
\begin{equation}
k_{i}^{j}=\left( 
\begin{array}{cc}
\frac{2\ddot{R}+f^{\prime }}{\sqrt{\dot{R}^{2}+f}} & -\frac{\left[
N_{+}^{\varphi }\left( R\right) +N_{-}^{\varphi }\left( R\right) \right] }{R}
\\ 
-\frac{\left[ N_{+}^{\varphi }\left( R\right) +N_{-}^{\varphi }\left(
R\right) \right] }{R} & \frac{2\sqrt{\dot{R}^{2}+f}}{R}%
\end{array}%
\right) .
\end{equation}%
Having $S_{i}^{j}$ diagonal implies that $N_{+}^{\varphi }\left( R\right)
+N_{-}^{\varphi }\left( R\right) =0$ which consequently admits $%
J_{+}+J_{-}=0.$ This in turn implies that the upper-shell and the
lower-shell are counterrotating, i. e. they spin in opposite directions.
This configuration is depicted in Fig. 1. (We should add that thin-shells in 
$3+1-$dimensions with counterrotating Kerr black holes have been constructed
before in \cite{Krisch}.) On the other hand, since $J^{2}$ appears in $f$,
it does not create any problem in having the other conditions satisfied.
Therefore 
\begin{equation}
-\frac{2\sqrt{\dot{R}^{2}+f}}{R}=8\pi G\sigma
\end{equation}%
\begin{equation}
\frac{2\ddot{R}+f^{\prime }}{\sqrt{\dot{R}^{2}+f}}=8\pi GP
\end{equation}%
are the only left Israel equations to be satisfied. With the assumption $%
J_{+}+J_{-}=0,$ therefore the problem becomes equivalent to the static (i.e.
non-rotating) case. At the equilibrium state where $R=R_{0}$ with $\dot{R}%
_{0}=\ddot{R}_{0}=0$ one gets%
\begin{equation}
\sigma _{0}=-\frac{2\sqrt{f_{0}}}{8\pi GR_{0}}
\end{equation}%
and%
\begin{equation}
P_{0}=\frac{f_{0}^{\prime }}{8\pi G\sqrt{f_{0}}}.
\end{equation}%
One observes easily that the energy conditions are not satisfied since $%
\sigma _{0}<0.$ The rest of the paper investigates the role of the angular
momentum on the stability of the TSW.

\section{Angular momentum and stability}

From Eq. (15) one finds%
\begin{equation}
\dot{R}^{2}+f-\frac{\left( 8\pi G\sigma \right) ^{2}R^{2}}{4}=0
\end{equation}%
which can be written as 
\begin{equation}
\dot{R}^{2}+V_{eff}\left( R\right) =0
\end{equation}%
with $V_{eff}=f-\frac{\left( 8\pi G\sigma \right) ^{2}R^{2}}{4}.$ Also, from
(15) and (16), it is a simple task to show that%
\begin{equation}
\sigma +P=-R\sigma ^{\prime }.
\end{equation}%
If we try to perturb the TSW from its equilibrium point, this relation must
be satisfied. To keep our analysis as general as possible, we consider $%
P=\xi \left( \sigma \right) $ in which $\xi $ is a well defined function of $%
\sigma $. One can show that $V_{eff}\left( R_{0}\right) =V_{eff}^{\prime
}\left( R_{0}\right) =0$ and therefore the first nonzero term of expansion
of $V_{eff}\left( R\right) $ about $R_{0}$ is $V_{eff}^{\prime \prime
}\left( R_{0}\right) $ given by%
\begin{equation}
V_{eff}^{\prime \prime }\left( R_{0}\right) =f_{0}^{\prime \prime }-\frac{%
f_{0}^{\prime 2}R_{0}^{2}+2\xi _{0}^{\prime }f_{0}\left(
2f_{0}-f_{0}^{\prime }R_{0}\right) }{2f_{0}R_{0}^{2}}
\end{equation}%
in which $\xi _{0}^{\prime }=\left. \frac{d\xi }{d\sigma }\right\vert
_{\sigma _{0}}$ and where quantities with a subindex $0$ implies they are
calculated at the throat $R=R_{0}.$ The equation of motion of the throat,
for small perturbation, becomes%
\begin{equation}
\dot{R}^{2}+\frac{V_{eff}^{\prime \prime }\left( R_{0}\right) }{2}\left(
R-R_{0}\right) ^{2}\tilde{=}0
\end{equation}%
or equivalently%
\begin{equation}
\ddot{x}+\frac{V_{eff}^{\prime \prime }\left( R_{0}\right) }{2}x\tilde{=}0
\end{equation}%
in which $x=R-R_{0}.$ Therefore for $V_{eff}^{\prime \prime }\left(
R_{0}\right) >0$ the motion of the throat is oscillatory with angular
frequency $\omega =\sqrt{\frac{V_{eff}^{\prime \prime }\left( R_{0}\right) }{%
2}}$ which implies that TSW is stable. On the other hand if $V_{eff}^{\prime
\prime }\left( R_{0}\right) <0$ motion is exponential type and hence the
throat upon the perturbation is unstable. In the following sections we shall
consider a specific EoS i.e. $\xi \left( \sigma \right) $ and for fixed
values of mass and $\ell ^{2}$ we study the effect of the angular momentum $%
J $ on stability of the wormhole.

\subsection{Linear Gas}

The EoS of a LG is given by $\frac{d\xi }{d\sigma }=\beta $ in which $\beta $
is a constant parameter. One observes that for the cases 
\begin{equation}
\frac{\left( 2f_{0}f_{0}^{\prime \prime }-f_{0}^{\prime 2}\right) R_{0}^{2}}{%
2f_{0}\left( 2f_{0}-f_{0}^{\prime }R_{0}\right) }<\beta \text{ \ \ \ \ \ for
\ \ \ }\frac{2f_{0}}{f_{0}^{\prime }}>R_{0}
\end{equation}%
and%
\begin{equation}
\frac{2f_{0}R_{0}^{2}f_{0}^{\prime \prime }-f_{0}^{\prime 2}R_{0}^{2}}{%
2f_{0}\left( 2f_{0}-f_{0}^{\prime }R_{0}\right) }>\beta \text{ \ \ \ \ \ for
\ \ \ \ }\frac{2f_{0}}{f_{0}^{\prime }}<R_{0}\text{ \ \ }
\end{equation}%
$V_{eff}^{\prime \prime }\left( R_{0}\right) >0$ and the equilibrium is
stable. In Fig. 2 we plot the stability region with respect to $\beta $ and $%
R_{0}.$ In the same figure the result for different $J$ are compared. As one
can see, increasing the value of $J$ increases the region of stability.
Therefore the RTSW supported by a perfect fluid with an EoS of the form of
LG, becomes more stable. Let's comment that, different EoS may also be
investigated but in this work LG is enough to show the contribution of the
angular momentum. 
%%%%%%%%%%%%%%%%%%%%%%%%%%%%%%%%%%%%%%%%%%%%%%%%%%%%%%%%%%%%%%%%%%%%%%%%%
\begin{figure}[tbp]
\includegraphics[width=70mm,scale=0.7]{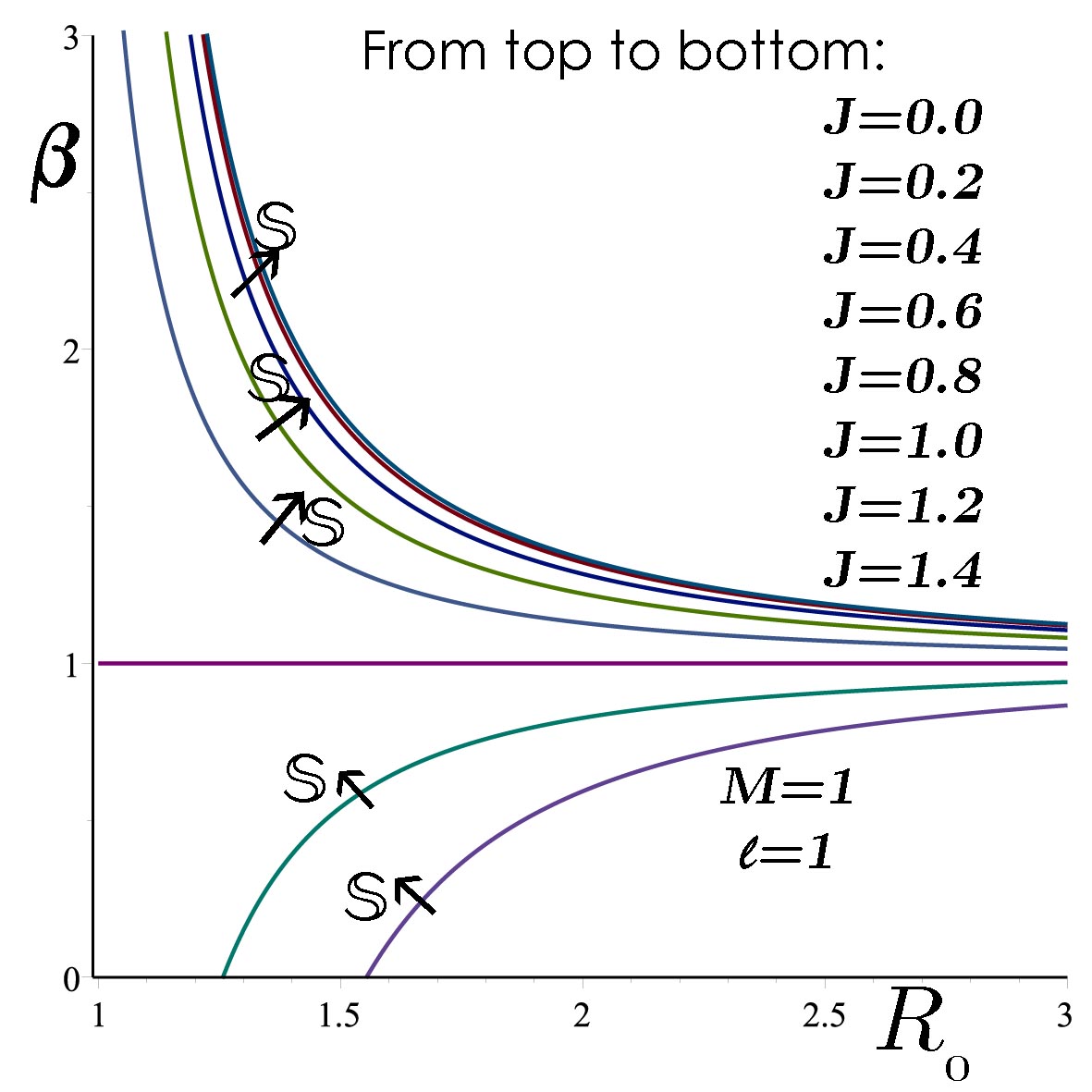}
\caption{A plot of regions of stability with respect to $R_{0}$ and $\protect%
\beta $ for $\ell =1.0$, $M=1.0$ and various values for the angular momentum 
$J.$ The value of $J$ is given from the top to the bottom as: $%
J=0.0,0.2,0.4,0.6,0.8,1.0,1.2$ and $1.4.$ We note that for $J=1.0$ the bulk
possesses a degenerate horizon (extremal black hole) and for $J>1$ the bulk
solution is not a black hole. The stability region for each case is shown
with an arrow. According to this figure, one concludes that the bigger
angular momentum, the more stable RTSW. }
\end{figure}
%%%%%%%%%%%%%%%%%%%%%%%%%%%%%%%%%%%%%%%%%%%%%%%%%%%%%%%%%%%%%%%%%%%%%%%%%%%%%

\subsection{Small velocity dependent perturbation}

In this section we apply the small velocity perturbation method which is
based on the assumption that the EoS of the fluid supporting the wormhole
after the perturbation is same as when the wormhole is at its equilibrium.
From this assumption one finds from Eqs. (15) and (16)%
\begin{equation}
\frac{P}{\sigma }=-\frac{Rf^{\prime }}{2f}
\end{equation}%
which yields the equation of motion for the throat after the small speed
perturbation as%
\begin{equation}
\frac{2\ddot{R}+f^{\prime }}{\dot{R}^{2}+f}=\frac{f^{\prime }}{f}.
\end{equation}%
Equivalently this amounts to 
\begin{equation}
\frac{d}{d\tau }\ln \left( \dot{R}^{2}+f\right) =\frac{d}{dR}\ln \left(
f\right)
\end{equation}%
which admits 
\begin{equation}
\dot{R}^{2}=\dot{R}_{0}^{2}f.
\end{equation}%
Using the explicit form of $f$ and upon integration yields 
\begin{equation}
\ln \left( \frac{2R^{2}\sqrt{f}+\ell \left( \frac{2R^{2}}{\ell ^{2}}%
-M\right) }{2R_{0}^{2}\sqrt{f_{0}}+\ell \left( \frac{2R_{0}^{2}}{\ell ^{2}}%
-M\right) }\right) =\frac{2\dot{R}_{0}}{\ell }\left( \tau -\tau _{0}\right) .
\end{equation}%
Perhaps it would be much easier to comment on $R$ if we could find a closed
form for it but even in this form one can see that the motion is not
oscillatory. This means that although the speed of the throat does not
increase dramatically with respect to $R$ the equilibrium is not stable.

\section{Conclusion}

For a TSW in 2+1-dimensions it is observed that by pasting two
counterrotating shells at the throat the off-diagonal elements of the
surface energy tensor $S_{i}^{j}$ (and related $K_{i}^{j}$) vanish. That is,
although $S_{\tau }^{\tau }$ and $S_{\psi }^{\psi }$ are not affected the
gluing procedure sets $S_{\tau }^{\psi }=S_{\psi }^{\tau }=0.$ This amounts
to the mutual cancellation of the upper and lower pressure components of the
fluid, leaving only the pressure component $S_{\psi }^{\psi }\neq 0.$

The effect of the rotation on the geometry, however, remains intact as it
depends on the square of the angular momentum ($\mathbf{J}^{2}$). Stability
of a counterrotating TSW for a gas of linear equation of state turns out to
make the TSW more stable. For very fast rotation the stability region grows
much larger in the parameter space. For a velocity dependent perturbation,
however, it is shown that TSW is no more stable. That is, perturbation of
the throat radius ($R_{0}$) that depends on initial speed ($\dot{R}_{0}\neq
0 $), no matter how small, doesn't return to the equilibrium radius $R_{0}$
again. Although our work is confined to the simple $2+1-$dimensional
spacetime it is our belief that similar behaviors are exhibited also by the
higher dimensional TSWs.

\end{document}